\documentclass[hyper, nofootinbib]{PoS}
\usepackage{comment}
\usepackage{amssymb,amsmath}
\usepackage{subfigure}
\usepackage{color}
\usepackage{graphicx}% Include figure files
\usepackage{enumerate}% For non-numerical items
\usepackage{bm}% bold math
\usepackage{slashed}

\usepackage{amssymb}
\usepackage{amsmath}
\usepackage{epsfig}

\title{Multichannel $1\rightarrow 2$ transition amplitudes in a finite volume}
\ShortTitle{Finite volume transition amplitude}

\author{Ra\'ul A. Brice\~no\\
	Thomas Jefferson National Accelerator Facility, 12000 Jefferson Avenue, Newport
  News, VA 23606, USA\\
        E-mail: \email{rbriceno@jlab.org}}

\author{Maxwell T. Hansen\\
Helmholtz Institute Mainz, Johannes Gutenberg-Universität, 55099 Mainz, Germany\\
E-mail: \email{mth28@uw.edu}}

\author{\speaker{Andr\'e Walker-Loud}\\
Thomas Jefferson National Accelerator Facility, 12000 Jefferson Avenue, Newport
  News, VA 23606, USA\\
  Department of Physics, The College of William and Mary, Williamsburg, Virginia 23187-8795, USA\\
        E-mail: \email{walkloud@wm.edu}}

\abstract{We derive a model-independent expression for finite-volume matrix elements. Specifically, we present a relativistic, non-perturbative analysis of the matrix element of an external current between a one-scalar in-state and a two-scalar out-state. Our result, which is valid for energies below higher-particle inelastic thresholds, generalizes the Lellouch-Luscher formula in two ways: we allow the external current to inject arbitrary momentum into the system and we allow for the final state to be composed an arbitrary number of strongly coupled two-particle states with arbitrary partial waves (including partial-wave mixing induced by the volume). We also illustrate how our general result can be applied to some key examples, such as heavy meson decays and meson photo production.
Finally, we point out complications that arise involving unstable resonance states, such as $B\rightarrow K^*\ell^+\ell^-$ when staggered or mixed-action/partially-quenched calculations are performed.}

\FullConference{The 32nd International Symposium on Lattice Field Theory,\\
		23-28 June, 2014\\
		Columbia University New York, NY}

\begin{document}

\section{Introduction}
One major application of lattice QCD (LQCD) is the determination of hadronic matrix elements of electroweak processes.
These calculations are necessary for interpreting the results of experimental searches for physics beyond the Standard Model as well as an improved understanding of Standard Model predictions.
In the last few years, we have seen the emergence of controlled calculations of processes involving multiple (two) hadrons in the initial and/or final state.
Unlike the single hadron ground state spectrum or matrix elements, there is no simple relation between the finite-volume (FV) matrix elements and the infinite volume ($\infty$V) transition amplitudes.
The complications arise from the fact that the multi-particle states can not be isolated asymptotically as single-particle non-interacting states, and the finite cubic box does not respect the full $O(3)$ spatial symmetry but only the cubic symmetry, so angular momentum is no a good quantum number.
For recent reviews of this topic, see Refs.~\cite{Briceno:2014tqa,Briceno:2014pka}.

In this talk, we review recent formalism that has been developed to study a wide class of $\textbf{1}\rightarrow\textbf{2}$ processes in the presence of an external current~\cite{Briceno:2014uqa}.
We generalize the original work by Lellouch and L\"{u}scher, which was derived for $K\rightarrow\pi\pi$ decays~\cite{Lellouch:2000pv} with specialized kinematics.  This pioneering work was extended to all elastic channels under the inelastic threshold~\cite{Lin:2001ek}.  For this S-wave kaon decay, in the rest-frame of the kaon, the partial wave mixing induced by the periodic cubic box is highly suppressed and was therefore not considered in these first works.
We were motivated to determine a \textit{master formula} derived with as few approximations as possible.
We present this master equation, relevant for pseudo-scalar particles where the final state is composed of any number of coupled channels.  We also demonstrate how partial-wave mixing is manifested in the determination of the infinite volume transition amplitudes from the finite volume matrix elements.%
\footnote{A similar work determining the volume modifications for $\gamma N\rightarrow \pi N$ was performed for restricted kinematics and no partial wave mixing, but with the inclusion of spin in Ref.~\cite{Agadjanov:2014kha}}

\section{Two-particles in a box: a review}

We begin with a brief review of the now well known quantization condition for two-particles in a finite volume with zero intrinsic spin, any number of open and coupled channels and any total momentum.  
This is a generalization of the well known two-particle L\"{u}scher formula~\cite{Luscher:1986pf,Luscher:1990ux} which was later extended to boosted two-particle systems~\cite{Rummukainen:1995vs,Kim:2005gf}.
The formula presented here was first obtained in Refs.~\cite{Hansen:2012tf,Briceno:2012yi}:%
\footnote{The discussion in this section follows closely that of Ref.~\cite{Kim:2005gf}.}
\begin{eqnarray}
\label{eq:QC}
{\det\left[\mathbb{K}(E_n)+ \left(\mathbb{F}^V(E_n)\right)^{-1}\right]}=0.
\end{eqnarray} 
In this equation, 
$\mathbb{K}$ is the two-particle K-matrix, which is directly related to the scattering amplitude, $\mathcal{M}$, via
%
%\begin{eqnarray}
%\label{eq:MKrel}
$\mathcal{M}^{-1}=\mathbb{K}^{-1}-i\mathbb{P}^{2}/2$, %\label{MMtildematrix}
%\end{eqnarray}
where $\mathbb{P}^2$ is the phase space factor, which is a diagonal matrix in the open channels and is defined in Ref.~\cite{Briceno:2014uqa}.
Although the K-matrix is diagonal in angular momentum space, the finite volume matrix, $\mathbb{F}^V$, is not. In general, this is a matrix that depends on the geometric and boundary conditions of the volume, as well as the energy  and momenta of the system.%
\footnote{The formula for $\mathbb{F}^V$ is generalized to systems with arbitrary intrinsic spin, in a volume that is a generic rectangular prism with twisted boundary conditions in Ref.~\cite{Briceno:2014oea}.} 
In a cubic box with periodic boundary conditions
\begin{eqnarray}
&& \left[\mathbb{F}^V_j\right]_{lm_l;l'm_{l'}}=-\frac{\xi_j}{8\pi P_{0,M}^*}\left[\sum_{l'',m''}\frac{(4\pi)^{3/2}}{k^{*{l''}}_{j,on}}c_{l''m''}^{\mathbf{d}}(k^{*2}_{j,on};{L}) 
 \int d\Omega~Y^*_{ lm_l}Y^*_{l''m''}Y_{l'm_{l'}}\right],
\label{eq:deltaGPBCs}
\end{eqnarray}
where the function $c^{\textbf{d}}_{lm}$ is defined as
\begin{eqnarray}
\label{eq:clm}
c^\mathbf{d}_{lm}(k^{*2}_j; {L})
=\frac{\sqrt{4\pi}}{\gamma L^3}\left(\frac{2\pi}{L}\right)^{l-2}\mathcal{Z}^\mathbf{d}_{lm}[1;(k^*_j {L}/2\pi)^2],
\hspace{1cm}
\mathcal{Z}^\mathbf{d}_{lm}[s;x^2]
= \sum_{\mathbf r \in \mathcal{P}_{\mathbf{d}}}\frac{|\mathbf{r}|^lY_{lm}(\mathbf{r})}{(r^2-x^2)^s}.\label{eq:clm}
\end{eqnarray} 
and $k^{*2}_{j,on}$ is the on-shell relative momenta of the two particles in the center of mass (c.o.m.) frame.
The determinant in Eq.~(\ref{eq:QC}) is evaluated over the inner product space of angular momentum and open two-body channels.%
\footnote{For a very nice application of this coupled channel formalism, see the work of the Hadron Spectrum Collaboration computing the scattering parameters in the $\pi K$-$K\eta$ system~\cite{Dudek:2014qha, Wilson:2014cna}.} 

It is useful to define the residue of the fully dressed two-particle propagator in finite volume in terms of the matrix $\mathbb{M}$ (the argument of the determinant in Eq.~(\ref{eq:QC}))
\begin{eqnarray}
\label{eq:mathbbM}
\mathbb{M}(P_{0,M})\equiv\mathbb{K}(P_{0,M})+ \left(\mathbb{F}^V(P_{0,M})\right)^{-1}.
\end{eqnarray}
The residue is then given by
\begin{eqnarray}
\mathcal{R}_{\Lambda_f,n_f}&=&{\mathcal M}^{-1\dagger}~\mathbb K~\left.\text{adj}[\mathbb{M}(P_{0,M})]~{\text{tr}\left[ \text{adj}[\mathbb{M}(P_{0,M})]\frac{\partial \mathbb{M}(P_{0,M})}{\partial P_{0,M}}\right]}^{-1}
~K~{\mathcal M}^{-1}
\right|_{P_{0,M}=E_{\Lambda_f,n_f}},
\label{eq:RLambdan}
\end{eqnarray}
where $\text{adj}[\mathbb{M}]=\mathbb{M}^{-1}\det[\mathbb{M}]$ is the adjugate of the matrix $\mathbb{M}$. $E_{\Lambda_f,n_f}$ is the $n_f$ eigenvalue of the $\Lambda_f$ irreducible representation (irrep) of the two-particle system.

%%%%%%%%%%%%%%%%%%%%%%%%%%%%%%%%%%%%%%%%%%%
\section{$\textbf{1}\rightarrow\textbf{2}$ matrix elements}
With the above formalism in hand, it is straightforward to arrive at the generalized Lellouch-L\"{u}scher result.
Let $\mathcal{A}_{\Lambda_f,n_f;\lambda_c}$ be the infinite volume transition amplitude for a $\textbf{1} \rightarrow\textbf{2}$ process under the presence of an external current, where $(\Lambda_f,n_f)$ denote the quantum numbers discussed above for the two-particle state and $\lambda_c$ labels the quantum numbers of the external current. 

Define $\left|\langle E_{\Lambda_f,n_f}\textbf{P}_f;L|\mathcal{J}_{\lambda_c}(0,\textbf{P}_f-\textbf{P}_i)| E_{\Lambda_i,0}\textbf{P}_i;L\rangle\right|$ as the finite volume matrix elements of the external current.
The resulting non-perturbative master equation relating these two quantities is
\begin{equation}
\left|\langle E_{\Lambda_f,n_f}\textbf{P}_f;L|\mathcal{J}_{\lambda_c}(0,\textbf{P}_f-\textbf{P}_i)| E_{\Lambda_i,0}\textbf{P}_i;L\rangle\right| =
\sqrt{ \frac{\mathcal{N}_i\mathcal{N}_f}{2E_{\Lambda_i,0}}}
\sqrt{
\left[\mathcal{A}^\dagger_{\Lambda_f,n_f;\lambda_c}~\mathcal R_{\Lambda_f,n_f}~\mathcal{A}_{\Lambda_f,n_f;\lambda_c}\right]
},
\label{eq:matJaieps}
\end{equation}
where $\mathcal{N}_i$($\mathcal{N}_f$) is the normalization of the initial(final) finite volume state, $| E_{\Lambda_i,0}\textbf{P}_i;L\rangle$ ($| E_{\Lambda_f,n_f}\textbf{P}_f;L\rangle$), which in Ref.~\cite{Briceno:2014uqa} was set to 1.
The subscript $\Lambda_i$ denotes the irrep of the single particle groundstate, which corresponds to $n_i=0$. Equation~\ref{eq:matJaieps}, which is the main result of this work, relates the finite volume matrix to, in general, an infinite number of transition amplitudes. 
$\mathcal A$ is understood as a column vector (and $\mathcal A^\dagger$ a row) in the spaced defined above (\emph{angular-momentum space} $\oplus$ \emph{open channel space}). 
Just as with the two-particle quantization condition, in practice one is required to performed a truncation over the contributing elements of the transition amplitude. For low-energies where the higher partial waves are suppressed, this is a justified approximation. 
For a detailed derivation, see Ref.~\cite{Briceno:2014uqa}.

\subsection{$K\rightarrow\pi\pi$}
In the original work by Lellouch and L\"{u}scher~\cite{Lellouch:2000pv}, the initial and final states were assumed to be at rest.
To recover this result, one first truncates the quantization condition, Eq.~(\ref{eq:QC}) to a system of S-wave interactions with suppression to all higher partial waves,
\begin{align}
\cot\delta_{S}+\cot\phi^\textbf{d}_{00}&=0,
\end{align}
where we have introduced the pseudo-phases $\phi^\textbf{d}_{lm}$ defined in terms of the $c^\mathbf{d}_{lm}$ functions, Eq.~(\ref{eq:clm}), 
\begin{eqnarray}
q^*_{\Lambda,n}\cot\phi^\textbf{d}_{lm}=-\frac{4\pi}{q^{*l}_{\Lambda,n}} c^\mathbf{d}_{lm}(q^{*2}_{\Lambda,n}; {L}).
\label{eq:philm}
\end{eqnarray} 
The ratio of the infinite transition amplitude to finite volume matrix element is then approximated
\begin{eqnarray}
\frac{|\mathcal{A}_{S,n_f}|^2}
{|\langle \pi\pi,E_{n_f}\textbf{P},\Lambda_f\mu_f;L|{\mathcal{J}}_{\text{weak}}(0,\textbf{0})| K,E_{K}\textbf{P};L\rangle|^2}=
\frac{16\pi E_{K}~E^{*}_{n_f}}{\mathcal{N}_i\mathcal{N}_f~q^{*}_{n_f}\xi}
\left.~\frac{\partial (\delta_S+\phi^\textbf{d}_{00})}
{\partial P_{0,M}}\right|
_{ P_{0,M}=E_{n_f}}.
\label{eq:LLfactor}
\end{eqnarray}
$E_{K}$ is the energy of the incoming kaon and the symmetry factor $\xi$ is equal to $1/2$ if the final two particles are identical and equal to 1 otherwise. Using Eq.~(\ref{eq:deltaGPBCs}), one can obtain the limit in which the final two particles do not interact,
\begin{align}
\frac{|\mathcal{A}_{S,n_f}|^2}
{|\langle \pi\pi,E_{n_f}\textbf{P},\Lambda_f\mu_f;L|{\mathcal{J}}_{\text{weak}}(0,\textbf{0})| K,E_{K}\textbf{P};L\rangle|^2}
\longrightarrow 
\frac{2E_K}{\mathcal{N}_i\mathcal{N}_f}~\mathcal{R}_{free}^{-1}\equiv \frac{2E_K}{\mathcal{N}_i\mathcal{N}_f}\frac{E_{n_f}^2}{\xi} L^3.
\label{eq:free}
\end{align}
In the limit of two identical final state pions with $E_{n_f} = E_K$, one recovers Ref.~\cite{Lellouch:2000pv}.

%%%%%%%%%%%%%%%%%%%%%%%%%%%%%%%%%%%%%%%%%%%%%%%%%
%\begin{table} 
%\begin{center}
%\begin{tabular}{c|c|c|c} 
%\hspace{.1cm}\textbf{d}\hspace{.1cm}& 
% \hspace{.1cm}$(00n)$\hspace{.1cm}& 
% \hspace{.1cm}$(nn0)$\hspace{.1cm}& 
% \hspace{.1cm}$(nnn)$\hspace{.1cm}
% \\\hline
%& $\alpha_{20,\mathbb{A}_1}^\mathbf{d} =\frac{2}{\sqrt{5}}$
%& $\alpha_{20,\mathbb{A}_1}^\mathbf{d} =-\frac{1}{\sqrt{5}},\hspace{.5cm}
%\alpha_{22,\mathbb{A}_1}^\mathbf{d} =-i\sqrt{\frac{6}{5}}$
%& $\alpha_{22,\mathbb{A}_1}^\mathbf{d} =-2i\sqrt{\frac{6}{5}}$
% \\ 
% & $\alpha_{20,\mathbb{E}}^\mathbf{d} =-\frac{1}{\sqrt{5}}$
%  & 
% $\alpha_{20,\mathbb{B}_1}^\mathbf{d} =-\frac{1}{\sqrt{5}},\hspace{.5cm}
%  \alpha_{22,\mathbb{B}_1}^\mathbf{d} =i\sqrt{\frac{6}{5}}$
% & $\alpha_{22,\mathbb{E}}^\mathbf{d} =i\sqrt{\frac{6}{5}}$\\
% && 
% $\alpha_{20,\mathbb{B}_2}^\mathbf{d} =\frac{2}{\sqrt{5}}$
%\vspace{.05cm}\\ \hline
%\end{tabular}
%\caption{Nonzero values of $\alpha_{20,\Lambda}^\mathbf{d}$ and $\alpha_{22,\Lambda}^\mathbf{d}$ for ${\mathbf{d}^2}\leq3$. For the $\mathbb{T}_1^-$ irrep of ${O}^D_h$, the $c^\mathbf{d}_{2m}$ vanish, therefore there is no need to define $\alpha_{2m,\Lambda}^\mathbf{d}$ for this irrep. 
%}
%\label{table:alphad}
%\end{center}
%\end{table} 
%%%%%%%%%%%%%%%%%%%%%%%%%%%%%%%%%%%%%%%%%%%%
%%%%%%%%%%%%%%%%%%Section%%%%%%%%%%%%%%%%%%%%%
%%%%%%%%%%%%%%%%%%%%%%%%%%%%%%%%%%%%%%%%%%%
\subsection{One nontrivial example: $\pi\gamma^*\rightarrow\pi\pi$ and $B\rightarrow\pi\pi+\ell\ell$}
Consider a final state with two degenerate pions in a P-wave.
Ignoring contamination from the F-wave, the spectrum will satisfy the equation
\begin{align}
\cot\delta_{P}+\cot\phi^\textbf{d}_{P}&=0,
\end{align}
where the pseudo phase $\phi^\textbf{d}_{P}$ can be written as a terms of the $\phi^\textbf{d}_{lm}$ defined in Eq.~(\ref{eq:philm})
\begin{align}
\cot\phi^\textbf{d}_{P}&\equiv\left(\cot\phi^\textbf{d}_{00}
+\alpha_{20,\Lambda}^\mathbf{d}\cot\phi^\textbf{d}_{20}
+\alpha_{22,\Lambda}^\mathbf{d}\cot\phi^\textbf{d}_{22} 
\right).
\label{eq:Ppseudo-phase}
\end{align}
where $\alpha_{20,\Lambda}^\mathbf{d}$ are constants and have been tabulated in Table~II of Ref.~\cite{Briceno:2014uqa} for irreps that have overlap with P-wave for symmetry groups corresponding to boosts $\sqrt{\textbf{d}^2}\leq2$.
The resulting ratio of infinite volume transition amplitude to finite volume matrix element is then 
\begin{eqnarray}
\frac{|\mathcal{A}_{\lambda_f;\lambda_c}|^2}
{|\langle \pi\pi,E_{n_f}\textbf{P}_f,\Lambda_f\mu_f;L|{\mathcal{J}}_{\lambda_c}(0,\textbf{P}_f-\textbf{P}_i)| \pi,E_{i}\textbf{P}_i;L\rangle|^2}
&=&
\frac{1}{\xi}\frac{16\pi E_{i}~E^{*}_{n_f}}{\mathcal{N}_i\mathcal{N}_f~q^{*}_{n_f}}
\left.~\frac{\partial (\delta_P+\phi^\textbf{d}_{P})}
{\partial P_{0,M}}\right|
_{ P_{0,M}=E_{n_f}},~~~
\label{eq:LLfactorP}
\end{eqnarray}
where $\lambda_f$ compactly encodes the labels associated with the final state that are explicitly discussed in Ref.~\cite{Briceno:2014uqa}. Unlike the scenario considered before, here the external current can inject arbitrary four-momenta and the symmetry factor $\xi$ has been set to one since the two particles must be distinguishable. This result is directly applicable for studying processes such as $\pi\gamma^*\rightarrow\pi\pi$ and $B\rightarrow\pi\pi+\ell\ell$.  For a decay process such as the one considered in Ref.~\cite{Horgan:2013pva}, $B\rightarrow\pi K+\ell\ell$, the final two hadrons are no longer degenerate and therefore S and P-wave will in general mix when the system is boosted. For these systems, one can still use Eq.~(\ref{eq:matJaieps}).
See Ref.~\cite{Briceno:2014uqa} for a detailed discussion of irreps and partial wave mixing in this system.

In this case, the non-interacting limit corresponds to Eq.~(\ref{eq:free}) with $\xi=1$.
For strongly interacting systems with rapidly varying phase shifts, the right hand side of Eq.~(\ref{eq:LLfactorP}) can largely deviate from it the free limit. 
Near an infinitely narrow resonances, the derivative of the phase shift will diverge.
Parameterizing the phase shift near $90^\circ$ as a relativistic Breit-Wigner
\begin{eqnarray}
\tan \delta=\frac{E^*~\Gamma_\rho}{(m_\rho^2-E^{*2})},
~~\Rightarrow~~
\left. 
\frac{\partial}{\partial E} \delta \right|_{E^*=m_\rho}=
\left. 
-\sin^2\delta~\frac{\partial}{\partial E} \cot\delta\right|_{E^*=m_\rho}
\approx 2\frac{E_\rho}{\Gamma_\rho~m_\rho},
\end{eqnarray}
where $E_\rho=\sqrt{P^2+m_\rho^2}$. 
As the width goes to zero, the derivative divergences, as expected. 
Ignoring contributions from the derivative of the pseudo-phase, which only diverges near the free-particle poles, one obtains the following result for the right hand side of Eq.~(\ref{eq:LLfactorP}) near a resonance mass with a narrow width
\begin{eqnarray}
\frac{|\mathcal{A}_{\lambda_f;\lambda_c}|^2}
{|\langle \pi\pi,E_{n_f}\textbf{P}_f,\Lambda_f\mu_f;L|{\mathcal{J}}_{\lambda_c}(0,\textbf{P}_f-\textbf{P}_i)| \pi,E_{i}\textbf{P}_i;L\rangle|^2}
&\longrightarrow&
\frac{1}{\xi}\frac{1}{\mathcal{N}_i\mathcal{N}_f}
2~E_i\frac{16 \pi E_\rho}{q^{*}_{\rho} ~\Gamma_\rho}
.~~~
\label{eq:LLfactorP2}
\end{eqnarray}
For a system with a resonance, one can parametrized the transition amplitude for a $X\rightarrow \pi\pi$, where $X$ is stable single particle state and the $\pi\pi$ system has been projected onto the $\rho$-channel, in the following form
\begin{eqnarray}
\mathcal{A}_{\lambda_f;\lambda_c}={F^{X\rightarrow\rho}_{\lambda_f;\lambda_c}(E^*,|\textbf{Q}|)}\frac{\sqrt{E^*\Gamma(E^*)}}{m_\rho^2-E_{cm}^2-i E^*\Gamma(E^*)}
\frac{1}{\sqrt{\xi}}
\sqrt{\frac{8\pi E^*}{q^*}}.
\end{eqnarray}
Note that although the numerator of the scattering amplitude in the presence of a resonance is proportional to $\Gamma(E^*)$, the transition amplitude is proportional to $\sqrt{\Gamma(E^*)}$. This can be understood by the fact that in a transition amplitude, only one of the external legs couples to the two-particle states, while in a scattering amplitude both incoming and outgoing states are composed of two-particle states. Near the resonance mass, the transition amplitude divergences inversely proportional to the square root of the resonance width
\begin{eqnarray}
\mathcal{A}_{\lambda_f;\lambda_c}
\longrightarrow
%{F^{X\rightarrow\rho}_{\lambda_f;\lambda_c}(m_\rho,|\textbf{Q}|)}\frac{\sqrt{\Gamma_\rho}}{-i m_\rho\Gamma_\rho}=
\frac{F^{X\rightarrow\rho}_{\lambda_f;\lambda_c}(m_\rho,|\textbf{Q}|)}
{-i\sqrt{\xi} \sqrt{q^*_\rho\Gamma_\rho/8\pi}},
\end{eqnarray}
where $F^{X\rightarrow\rho}_{\lambda_f;\lambda_c}(m_\rho,|\textbf{Q}|)$ is the $X\rightarrow\rho$ transition amplitude evaluated at $E^*=m_\rho$.

In this limit, one finds that the finite volume matrix element calculated via lattice QCD is equal to the infinite volume $X\rightarrow\rho$ transition amplitude up to the standard normalization of the states,
\begin{eqnarray}
|\langle \pi\pi,E_{n_f}\textbf{P}_f,\Lambda_f\mu_f;L|{\mathcal{J}}_{\lambda_c}(0,\textbf{P}_f-\textbf{P}_i)| \pi,E_{i}\textbf{P}_i;L\rangle|^2
\longrightarrow
{|F^{X\rightarrow\rho}_{\lambda_f;\lambda_c}(m_\rho,|\textbf{Q}|)|^2} 
\frac{\mathcal{N}_i\mathcal{N}_f}
{2E_i~2 E_\rho}
.~~~
\label{eq:pi_to_Rho}
\end{eqnarray}
It is important to emphasize that this approximation only holds when one knows that a resonance is very narrow ($\Gamma_R/m_R\ll 1$) and the energy level determined corresponds to the resonances mass, up to small correction that scale with the width. To reliably asses the validity of this approximation, one must first determine the phase shift as a function of the energy using Eq.~(\ref{eq:QC}). 

%%%%%%%%%%%%%%%%%%%%%%%%%%%%%%%%%%%%%%%%%%%%%%%%%%%%%%%

\subsection{Comment on the calculation of $B\rightarrow K^* \ell^+\ell^-$ with staggered fermions}

Promising observables to search for new physics are the rare decays $B\rightarrow K^*\ell^+\ell^-$ and $B_s\rightarrow\phi\ell^+\ell^-$.  These processes were recently addressed with lattice QCD calculations~\cite{Horgan:2013hoa,Horgan:2013pva} utilizing staggered fermions.
Unfortunately, due to the resonant nature of the final state, the use of staggered fermions (or mixed-action/partially-quenched calculations) present a technical challenge which likely invalidate the results for direct comparison with physical transition amplitudes.
The problem is that the final state $K^*$ can decay strongly to the two-particle $K\pi$ system, including disconnected quark-flow diagrams.
In a calculational scheme that does not completely respect unitarity, such as the rooted staggered formulation (or mixed-action/partially-quenched schemes), these strong decays will contain so called \textit{hair-pin} interactions with unphysical double poles~\cite{Bernard:1992mk,Bernard:1993sv,Sharpe:1997by}.
These hair-pin interactions in the s-channel diagrams were shown to invalidate the know relation between two-particle energy levels in a finite volume and the infinite volume scattering phase shift~\cite{Bernard:1995ez}, as the optical theorem is not satisfied even below the inelastic threshold.
Scattering in maximally stretched isospin channels is protected from these issues in the elastic region, but suffers the same problem above inelastic thresholds~\cite{Chen:2005ab}.
Consequently, as the $K^*$ final state in the above mentioned calculations~\cite{Horgan:2013hoa,Horgan:2013pva} can strongly decay, we do not know formal relation between the finite-volume matrix elements which were computed and the corresponding infinite volume transition amplitudes, nor do we know how to quantify the error made in the calculations.
For these reasons, processes such as these require a numerical formulation which respects unitarity, at least at this time.

\section{Conclusion and discussion}
In this work we have reviewed recent work that allows for the determination of transition amplitudes for $\textbf{1}\rightarrow\textbf{2}$ processes where the individual particles have no intrinsic spin.
This is relevant for meson photo production and heavy meson decays for example.
We have found a universal equation relating finite volume matrix elements and infinite volume transition amplitudes, Eq.~(\ref{eq:matJaieps}). We have reviewed two simple examples on its implementation, $K\rightarrow\pi\pi$ and $X\rightarrow\pi\pi$ (where $X$ is a stable single particle state and final state is projected onto a P-wave). Furthermore, we discuss the free limit and the narrow width limit of our results.  

The steps needed to reliably determine $\textbf{1}\rightarrow\textbf{2}$ transition amplitudes via lattice QCD are:

1. \emph{Determine two-particle spectrum:} For systems involving resonances this requires utilizing a wide range of operators to maximally disentangle the energy levels. For example, in the $\pi K$-$K\eta$ calculation by Wilson et. al \cite{Dudek:2014qha, Wilson:2014cna}, this required using ``$q\bar{q}$-\emph{like}'', ``$\pi K$-\emph{like}'', and ``$K\eta$-\emph{like}'' operators which requires the evaluation disconnected diagrams.

2. \emph{Plug the finite volume spectrum into Eq.~(\ref{eq:QC}) to determine the K-matrix} 

3. \emph{Parametrize K-matrix:} By parametrizing the K-matrix one can then analytically take its derivative as a function of the energy of the system and consequently determine the two-particle residue using Eq.~(\ref{eq:RLambdan}).

4. \emph{Evaluate finite volume matrix element via lattice QCD}

5. \emph{Determine the $\textbf{1}\rightarrow\textbf{2}$ transition amplitude using Eq.~(\ref{eq:matJaieps}).}

\section*{Acknowledgements}
R.B. and A.W-L. acknowledge support from the U.S. Department of Energy (DOE) contract DE-AC05-06OR23177, under which Jefferson Science Associates, LLC, manages and operates the Jefferson Lab.  The work of A.W-L. was also supported by the U.S. DOE Early Career Award contract DE-SC0012180.
The work of M.T.H. was supported in part by DOE grant No. DE-FG02-96ER40956.


\begin{thebibliography}{99}

\bibitem{Briceno:2014tqa} 
  R.~A.~Briceño, Z.~Davoudi and T.~C.~Luu,
  %``Nuclear Reactions from Lattice QCD,''
  arXiv:1406.5673 [hep-lat].
  %%CITATION = ARXIV:1406.5673;%%
\bibitem{Briceno:2014pka} 
  R.~A.~Briceño,
  %``Few-body physics,''
  arXiv:1411.6944 [hep-lat].
  %%CITATION = ARXIV:1411.6944;%%
\bibitem{Briceno:2014uqa} 
  R.~A.~Briceño, M.~T.~Hansen and A.~Walker-Loud,
  %``Multichannel one-to-two transition form factors in a finite volume,''
  arXiv:1406.5965 [hep-lat].
  %%CITATION = ARXIV:1406.5965;%%
\bibitem{Lellouch:2000pv} 
  L.~Lellouch and M.~Luscher,
  %``Weak transition matrix elements from finite volume correlation functions,''
  Commun.\ Math.\ Phys.\  {\bf 219}, 31 (2001)
  [hep-lat/0003023].
  %%CITATION = HEP-LAT/0003023;%%
\bibitem{Lin:2001ek} 
  C.~J.~D.~Lin, G.~Martinelli, C.~T.~Sachrajda and M.~Testa,
  %``K --> pi pi decays in a finite volume,''
  Nucl.\ Phys.\ B {\bf 619}, 467 (2001)
  [hep-lat/0104006].
  %%CITATION = HEP-LAT/0104006;%%
\bibitem{Agadjanov:2014kha} 
  A.~Agadjanov, V.~Bernard, U.~G.~Meißner and A.~Rusetsky,
  %``A framework for the calculation of the $?N?^?$ transition form factors on the lattice,''
  Nucl.\ Phys.\ B {\bf 886}, 1199 (2014)
  [arXiv:1405.3476 [hep-lat]].
  %%CITATION = ARXIV:1405.3476;%%

\bibitem{Luscher:1986pf} 
  M.~Luscher,
  %``Volume Dependence of the Energy Spectrum in Massive Quantum Field Theories. 2. Scattering States,''
  Commun.\ Math.\ Phys.\  {\bf 105}, 153 (1986).
  %%CITATION = CMPHA,105,153;%%
\bibitem{Luscher:1990ux} 
  M.~Luscher,
  %``Two particle states on a torus and their relation to the scattering matrix,''
  Nucl.\ Phys.\ B {\bf 354}, 531 (1991).
  %%CITATION = NUPHA,B354,531;%%
\bibitem{Rummukainen:1995vs} 
  K.~Rummukainen and S.~A.~Gottlieb,
  %``Resonance scattering phase shifts on a nonrest frame lattice,''
  Nucl.\ Phys.\ B {\bf 450}, 397 (1995)
  [hep-lat/9503028].
  %%CITATION = HEP-LAT/9503028;%%
\bibitem{Kim:2005gf} 
  C.~h.~Kim, C.~T.~Sachrajda and S.~R.~Sharpe,
  %``Finite-volume effects for two-hadron states in moving frames,''
  Nucl.\ Phys.\ B {\bf 727}, 218 (2005)
  [hep-lat/0507006].
  %%CITATION = HEP-LAT/0507006;%%

\bibitem{Hansen:2012tf} 
  M.~T.~Hansen and S.~R.~Sharpe,
  %``Multiple-channel generalization of Lellouch-Luscher formula,''
  Phys.\ Rev.\ D {\bf 86}, 016007 (2012)
  [arXiv:1204.0826 [hep-lat]].
  %%CITATION = ARXIV:1204.0826;%%
\bibitem{Briceno:2012yi} 
  R.~A.~Briceno and Z.~Davoudi,
  %``Moving multichannel systems in a finite volume with application to proton-proton fusion,''
  Phys.\ Rev.\ D {\bf 88}, no. 9, 094507 (2013)
  [arXiv:1204.1110 [hep-lat]].
  %%CITATION = ARXIV:1204.1110;%%
\bibitem{Briceno:2014oea} 
  R.~A.~Briceno,
  %``Two-particle multichannel systems in a finite volume with arbitrary spin,''
  Phys.\ Rev.\ D {\bf 89}, no. 7, 074507 (2014)
  [arXiv:1401.3312 [hep-lat]].
  %%CITATION = ARXIV:1401.3312;%%
\bibitem{Dudek:2014qha} 
  J.~J.~Dudek {\it et al.}  [Hadron Spectrum Collaboration],
  %``Resonances in coupled $\pi K -\eta K$ scattering from quantum chromodynamics,''
  Phys.\ Rev.\ Lett.\  {\bf 113}, no. 18, 182001 (2014)
  [arXiv:1406.4158 [hep-ph]].
  %%CITATION = ARXIV:1406.4158;%%
\bibitem{Wilson:2014cna} 
  D.~J.~Wilson, J.~J.~Dudek, R.~G.~Edwards and C.~E.~Thomas,
  %``Resonances in coupled $\pi K, \eta K$ scattering from lattice QCD,''
  arXiv:1411.2004 [hep-ph].
  %%CITATION = ARXIV:1411.2004;%%


\bibitem{Horgan:2013hoa} 
  R.~R.~Horgan, Z.~Liu, S.~Meinel and M.~Wingate,
  %``Lattice QCD calculation of form factors describing the rare decays $B \to K^* \ell^+ \ell^-$ and $B_s \to \phi \ell^+ \ell^-$,''
  Phys.\ Rev.\ D {\bf 89}, no. 9, 094501 (2014)
  [arXiv:1310.3722 [hep-lat]].
  %%CITATION = ARXIV:1310.3722;%%
\bibitem{Horgan:2013pva} 
  R.~R.~Horgan, Z.~Liu, S.~Meinel and M.~Wingate,
  %``Calculation of $B^0 \to K^{*0} \mu^+ \mu^-$ and $B_s^0 \to \phi \mu^+ \mu^-$ observables using form factors from lattice QCD,''
  Phys.\ Rev.\ Lett.\  {\bf 112}, 212003 (2014)
  [arXiv:1310.3887 [hep-ph]].
  %%CITATION = ARXIV:1310.3887;%%
\bibitem{Bernard:1992mk} 
  C.~W.~Bernard and M.~F.~L.~Golterman,
  %``Chiral perturbation theory for the quenched approximation of QCD,''
  Phys.\ Rev.\ D {\bf 46}, 853 (1992)
  [hep-lat/9204007].
  %%CITATION = HEP-LAT/9204007;%%
\bibitem{Bernard:1993sv} 
  C.~W.~Bernard and M.~F.~L.~Golterman,
  %``Partially quenched gauge theories and an application to staggered fermions,''
  Phys.\ Rev.\ D {\bf 49}, 486 (1994)
  [hep-lat/9306005].
  %%CITATION = HEP-LAT/9306005;%%
\bibitem{Sharpe:1997by} 
  S.~R.~Sharpe,
  %``Enhanced chiral logarithms in partially quenched QCD,''
  Phys.\ Rev.\ D {\bf 56}, 7052 (1997)
  [Erratum-ibid.\ D {\bf 62}, 099901 (2000)]
  [hep-lat/9707018].
  %%CITATION = HEP-LAT/9707018;%%
\bibitem{Bernard:1995ez} 
  C.~W.~Bernard and M.~F.~L.~Golterman,
  %``Finite volume two pion energies and scattering in the quenched approximation,''
  Phys.\ Rev.\ D {\bf 53}, 476 (1996)
  [hep-lat/9507004].
  %%CITATION = HEP-LAT/9507004;%%
\bibitem{Chen:2005ab} 
  J.~W.~Chen, D.~O'Connell, R.~S.~Van de Water and A.~Walker-Loud,
  %``Ginsparg-Wilson pions scattering on a staggered sea,''
  Phys.\ Rev.\ D {\bf 73}, 074510 (2006)
  [hep-lat/0510024].
  %%CITATION = HEP-LAT/0510024;%%

\end{thebibliography}
\end{document}